\documentclass[aps,prd,reprint,superscriptaddress, showpacs]{revtex4-2}

\usepackage{graphicx}
\usepackage{amsmath}
\usepackage{color}
\begin{document}


\title{Search for an exotic parity-odd spin- and velocity-dependent interaction using a magnetic force microscope}

\author{Xiaofang Ren}
\affiliation{MOE Key Laboratory of Fundamental Quantities Measurement, Hubei Key Laboratory of Gravitation and Quantum Physics, School of Physics, Huazhong University of Science and Technology, Wuhan 430074, China}

\author{Jianbo Wang}
\affiliation{MOE Key Laboratory of Fundamental Quantities Measurement, Hubei Key Laboratory of Gravitation and Quantum Physics, School of Physics, Huazhong University of Science and Technology, Wuhan 430074, China}

\author{Rui Luo}
\affiliation{MOE Key Laboratory of Fundamental Quantities Measurement, Hubei Key Laboratory of Gravitation and Quantum Physics, School of Physics, Huazhong University of Science and Technology, Wuhan 430074, China}
\author{Lichang Yin}
\affiliation{Shenyang National Laboratory for Materials Science, Institute of Metal Research, Chinese Academy of Sciences, Shenyang 110016, China}
\affiliation{Department of Physics and Electronic Information, Huaibei Normal University, Anhui, Huaibei 235000, China}

\author{Jihua Ding}
\affiliation{MOE Key Laboratory of Fundamental Quantities Measurement, Hubei Key Laboratory of Gravitation and Quantum Physics, School of Physics, Huazhong University of Science and Technology, Wuhan 430074, China}

\author{Ge Zeng}
\affiliation{School of Physics, Huazhong University of Science and Technology, Wuhan 430074, China}

\author{Pengshun Luo}
\email[E-mail: ]{pluo2009@hust.edu.cn}
\affiliation{MOE Key Laboratory of Fundamental Quantities Measurement, Hubei Key Laboratory of Gravitation and Quantum Physics, School of Physics, Huazhong University of Science and Technology, Wuhan 430074, China}

\date{\today}

\begin{abstract}
Exotic spin-dependent interactions may be generated by exchanging hypothetical bosons that have been proposed to solve some mysteries in physics by theories beyond the standard model of particle physics. The search for such interactions can be conducted by tabletop scale experiments using high precision measurement techniques. Here we report an experiment to explore the parity-odd interaction between moving polarized electrons and unpolarized nucleons using a magnetic force microscope. The polarized electrons are provided by the magnetic tip at the end of a silicon cantilever, and their polarizations are approximately magnetized in the plane of the magnetic coating on the tip. A periodic structure with alternative gold and silicon dioxide stripes provides unpolarized nucleons with periodic number density modulation. The exotic forces are expected to change the oscillation amplitude of the cantilever which is measured by a fiber laser interferometer. Data has been taken by scanning the tip over the nucleon source structure at constant separation, and no exotic signal related to the density modulation has been observed. Thus, the experiment sets a limit on the electron-nucleon coupling constant, $g_A^eg_V^N\leq 9\times 10^{-15}$ for 15 $\mu$m $\le \lambda \le$ 180 $\mu$m, using a direct force measurement method.
\end{abstract}


\maketitle
\section{Introduction}	

The experimental search for the exotic spin-dependent interactions has received substantial attention in the last decade\cite{Adelberger2009, Ni2010, Safronova2018}. These interactions may mediated by hypothetical bosons such as axions\cite{Peccei1977,Weinberg1978, Wilczek1978,Kim1979}, familons\cite{Wilczek1982, GELMINI1983}, Majorons\cite{GELMINI1981,CHIKASHIGE1981}, arions\cite{Anselm1982}, paraphotons\cite{Okun1982, Holdom1986, Dobrescu2005}, and new $Z'$ bosons\cite{BOUCHIAT1983, Appelquist2003, Dzuba2017}.  These new bosons are introduced in efforts to resolve some puzzles in physics such as the strong CP problem\cite{Peccei1977, Weinberg1978, Wilczek1978}, the nature of dark matter and dark energy\cite{Bertone2018,Frieman2008},  and the hierarchy problem \cite{Graham2015}. Under the framework of quantum field theory, Dobrescu and Mocioiu derived sixteen types of interaction potentials generated by any generic spin-0 or spin-1 boson exchange assuming rotational invariance\cite{Bogdan2006}. These potentials were re-derived and classified according to the types of physical couplings recently by Fadeev $et$ $al.$ \cite{Fadeev2019}.  Except for the Yukawa-type interaction, fifteen of them are spin dependent and generated through pseudoscalar, vector or axial-vector couplings.

Among them, extensive searches \cite{ Wineland1991, Venema1992, Youdin1996, Ni1999, Heckel2006, Hammond2007, Petukhov2010, Hoedl2011,Tullney2013, Bulatowicz2013, Terrano2015, Lee2018, Stadnik2018, Rong2018b, Chui1993, Glenday2008, Vasilakis2009, Heckel2013, Kotler2015, Luo2017, Rong2018, Ficek2018, Almasi2020} have been conducted on the axion-mediated spin-mass  or spin-spin exotic interactions  that were initially proposed by Moody and Wilczek\cite{Moody1984}.  Experimental searches have also been performed on the spin- and velocity- dependent exotic interactions\cite{Heckel2006, Heckel2008, Adelberger2013, Piegsa2012, Kim2018, Ficek2018,Ding2020,  Yan2013, Yan2015, Dzuba2017, Parnell2020}.   In this paper, we explore one of them, which is parity-odd as given by
\begin{equation}
{V_{12+13}} =g_A^eg_V^N\frac{\hbar}{4\pi}(\hat{\sigma}\cdot\vec{v})\frac{1}{r}e^{-r/\lambda},
\label{eq1}
\end{equation}
where $\hbar$ is the reduced Planck's constant, $g_A^e$ is the axial-vector coupling constant to electrons, $g_V^N$ is the vector coupling constant to nucleons, $\hat{\sigma}$ is the Pauli vector of electron spin, $\vec{v}$ is the relative velocity between electron and nucleon, and $\lambda = \hbar/m_bc$ is the interaction range with boson mass $m_b$.  This potential can be generated by the exchange of a massive spin-1 boson $Z'$  in a Lorentz-invariant theory described by the Lagrangian $\mathcal{L}_{Z'} = Z'_\mu\bar{\psi}\gamma^\mu(g_\psi^V + \gamma_5g_\psi^A)\psi$\cite{Bogdan2006, Fadeev2019}, here $\psi$ denotes the fermion field.  For simplicity, we assume that $g_V^p=g_V^n=g_V^N$,  where $p$ and $n$ denote proton and neutron, respectively.

The above potential can be treated as $V = \gamma\hbar \hat\sigma\cdot\vec B_{eff}$, and then we can see that the spin $\hat{\sigma}$ can be used to sense the effective magnetic field $\vec B_{eff}$ produced by the interaction. Atomic magnetometer operating in the spin-exchange relaxation-free (SERF) regime has been used to detect the $\vec B_{eff}$ produced by a unpolarized bismuth germanate insulator, and the experiment sets the strongest constraints on the coupling constant between electrons and nucleons for $\lambda >1 \times 10^{-4} \ m$ \cite{ Kim2019}. The measurements of neutron spin rotation or spin relaxation rate of polarized $^3$He gas have also been used to constrain the possible new interaction between polarized neutrons and unpolarized matter\cite{Yan2013, Yan2015}.  The exchange of vector bosons between electrons and nucleons induces parity-nonconserving effects (PNC) in atoms and molecules, which was used to constrain the parity violating vector axial-vector nucleon-electron and nucleon-proton interactions\cite{Dzuba2017} at the nanometer range.  Recently, the exotic interaction between the neutron and the matter of the Earth was sought using a spin-echo based interferometry technique  \cite{Parnell2020}.  On the other hand, spin polarized torsion pendulum is one of the most sensitive technique to search for the new interaction at centimeter scale by macroscopic torque measurement\cite{Heckel2006,Heckel2008,  Adelberger2013}. Here we use a sensitive cantilever with a magnetic tip to search for the interaction between polarized electrons and unpolarized nucleons at the micrometer scale by direct force measurement.

\begin{center}
  \begin{figure}
	\includegraphics[width=8.5cm]{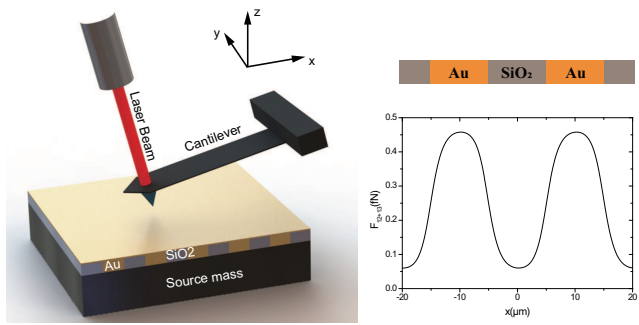}
	\caption{(a) Schematic drawing of the experiment. Dimensions are not in scale. (b) The lateral position ($x$) dependence of the exotic force amplitude calculated at a distance of 730 nm with $\lambda$ =1 $\mu$m, $g_A^eg_V^N=2\times 10^{-13}$.}
	\label{Schematic}
\end{figure}
\end{center}

\section{Experimental scheme}	
The experiment is schematically shown in Fig.~\ref{Schematic}(a). A cantilever with a magnetic tip is used as a force sensor to measure the force between the spin-polarized magnetic tip and a unpolarized nucleon source at the micrometer range. The magnetic tip is coated with hard magnetic CoCr alloy to provide spin-polarized electrons. The tip was magnetized in the perpendicular direction to the cantilever and the electron polarizations are approximately in the plane of the magnetic coating. As the $V_{12+13}$ potential is proportional to $(\hat{\sigma}\cdot\vec{v})$, the relative velocity is chosen to be along the polarization direction in order to maximize the exotic force. This is fulfilled by driving the cantilever vibrating perpendicularly. The source mass is made of alternative high density (gold) and low density (silicon dioxide) materials to create a nucleon density modulation. When the tip is oscillating over the source mass,  the vibration of the cantilever is reinforced or weakened by the exotic interaction, and the oscillation amplitude changes depending on the density of the material underneath the tip. Therefore, a periodic signal correlated with the density modulation would be expected if the exotic interaction is detectable. This design helps to separate the signal of interest from the spurious signals, such as the electrostatic force and Casimir force.

The cantilever is driven by a piezo actuator at given amplitude and frequency, and its displacement is measured by a home-made fiber laser interferometer. The exotic interaction acts as an additional damping (or exciting) force proportional to the velocity of the cantilever. The equation of motion of the cantilever can be written as
\begin{equation}
m\ddot{z}+m\gamma\dot{z}+kz=F_d\cos(\omega_d t)+f_s\dot{z},
\label{eq2}
\end{equation}
where $m$, $k$, and $\gamma$ are the effective mass, spring constant and damping factor of the cantilever, respectively. $F_d$ is the external driving force amplitude of the piezo, and $\omega_d$ is the driving angular frequency. The $f_s=F_s/v$ is the spin-dependent exotic force ($F_s$) divided by velocity. In this study, the cantilever is constantly driven at its resonance frequency, and the cantilever's vibrational amplitude is given by
\begin{equation}
z_{amp}\approx z_{far}\left(1+\frac{Q\omega_0f_s}{k}\right) ,
\label{eq3}
\end{equation}
where $z_{far}=QF_d/k$ is the vibration amplitude when the magnetic tip is free from any extra interaction, which can be determined when the magnetic tip is far away from the source mass. $Q=\omega_0/\gamma$ is the quality factor of the cantilever, $\omega_0$ is its resonance angular frequency. To avoid the interference of constant background signals, we search for the exotic interaction by looking for the amplitude variation when the magnetic tip scans over different areas of nucleon density.

\section{Experimental details}	

\subsection{Magnetic force probe}	

\begin{figure*}
	\includegraphics {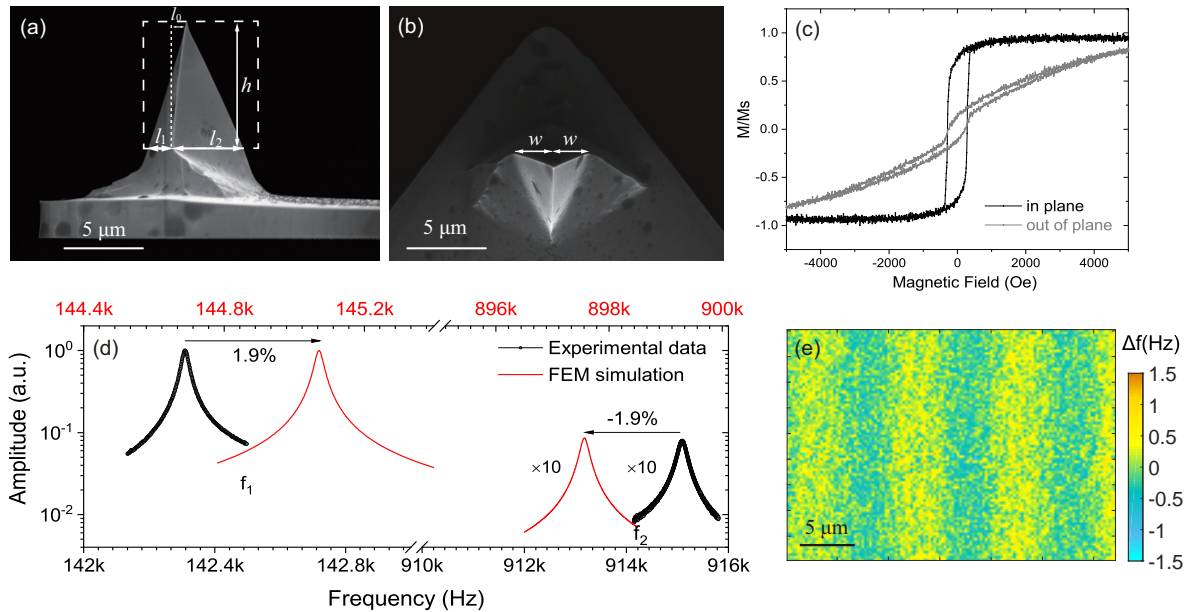}
	\caption{(a), (b) SEM images of the magnetic tip. (c) The hysteresis of the CoCr coating on the silicon substrate measured with magnetic field applied in plane and out of plane, respectively. (d) The measured first and second resonances of the cantilever in comparison to the results from finite element simulation. (e) MFM image ($\Delta f$) of the periodically magnetized hard-disk plate taken with the magnetic tip. }
	\label{MFM_Tip}
\end{figure*}

A commercially available magnetic force probe was used in this experiment\cite{MFMtip00,Lohau1999,Babcock1994}. The cantilever is made of single crystalline silicon with  dimension of 121.90 $\mu$m $\times$ 34.77 $\mu$m $\times$ 1.62 $\mu$m as measured by scanning electron microscopy (SEM). The eigenfrequencies of the first two bending modes were measured to be $f_1=142309$ Hz and $f_2  =915089$ Hz, respectively. We determined the effective spring constant of the cantilever by matching the two measured eigenfrequencies to the values calculated by finite element analysis. The frequency match is achieved with a relative uncertainty of 1.9\% by only tuning the cantilever thickness to 1.415 $\mu$m (Fig.~\ref{MFM_Tip}(d)). The thickness is close to the measured value after subtracting the coating layer. The difference in resonance frequency between the experiment and the simulation should be due to the fact that the coating is not considered in the simulation, because of the lack of affirmative property inputs, and its much smaller thickness compared to the silicon cantilever. The spring constant is derived from the simulation to be 3.6(4)  N/m, and the main uncertainty is contributed from the uncertain location of the laser spot on the cantilever.

The magnetic tip has a pyramid shape, and is coated by a layer of hard magnetic CoCr alloy film with a Co:Cr ratio of 80:20. The nominal thickness of the coating is 50 nm. The magnetic properties of the coating layer were evaluated by a vibration sample magnetometer (VSM). Since the tip is too tiny to contribute a detectable signal, we instead measured the magnetization curve  of the silicon substrate that holds the tip. As the CoCr alloy was deposited on the silicon substrate and tip simultaneously, we assume that the coatings have the same quality. Figure~\ref{MFM_Tip}(c) shows the in-plane and out-of-plane magnetization curves which is qualitatively consistent with previous measurements\cite{Hopkins1996, Jaafar2008}. The results indicate that the easy axis is in the plane and the in-plane hysteresis has a squareness ($M_r/M_s$) of 0.76, so that we would expect that the magnetic moments of the magnetized tip would domainantly stay parallel to the tip surface and point along the applied field direction. The  remnant magnetization per area is calculate from the measurement to be  $M_r^a = 0.034$ A, which is used to calculate the exotic forces later.

In order to qualitatively verify the magnetism of the tip, we have used the tip to image a hard-disk plate with magnetic force microscopy (MFM). The hard-disk plate was magnetized to periodic patterns using a home-made magnetic writing device. The period of the pattern is $\sim$ 10 $\mu$m.  In the MFM imaging, the topography is measured in the trace scan and then the resonance frequency shift ($\Delta f$) is measured in the retrace scan at a lifted height. Figure~\ref{MFM_Tip}(e) shows the frequency shift image taken at a lifted height of 150 nm. The period of the image is consistent with the expected value, which qualitatively verify the magnetism of the tip. The images have been taken both before and after the experimental runs, and we observed no significant change between those images.

\subsection{ Nucleon source structure }	

\begin{figure}
	\includegraphics [width=8cm]{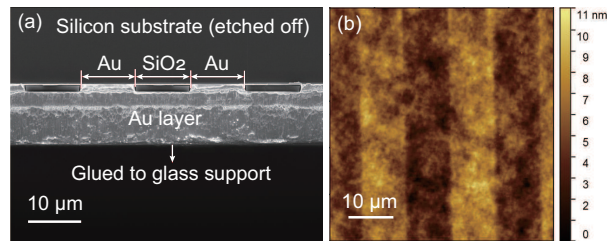}
	\caption{(a) The cross section of the density modulation source structure. (b) The AFM topography image of the source structure surface. }
	\label{Sample}
\end{figure}

A periodic structure with alternative high density (gold) and low density  (silicon dioxide) materials is used to provide the source of nucleons. The density modulation structure was fabricated on a silicon wafer topped with a 200 nm thick SiO$_{2}$ layer. A 190 nm thick gold film was first evaporated on the oxide as a conductive layer, and then another 1.01 $\mu$m thick SiO$_{2}$ layer was deposited on the gold layer by plasma enhanced chemical vapor deposition (PECVD), which serves as the low-density material. Afterwards, the SiO$_{2}$ layer was patterned to parallel trenches with a period of  20.10 $\mu$m using optical lithography and reactive ion etching (RIE). Each trench has a width of $\sim$ 10 $\mu$m and a depth of 1.01 $\mu$m. The trench was then filled with gold by electroplating, which is used as the high nucleon density material. Finally, the structure was glued on a glass substrate, and then the silicon wafer was removed by alkali etching (20 \% KOH with addition of hydroxylamine) followed by HF acid etching of the oxide layer. Figure ~\ref{Sample} presents a SEM image of the cross section of the nucleon source structure, the dimensions of the structure are listed in Table~\ref{parameter}. The surface roughness is around 4 nm as measured with atomic force microscopy (AFM), and the averaged periodic corrugation is $\sim$ 2 nm.

{\begin{table}
\caption{Mean values and uncertainties of the main experimental parameters.}
  \centering
\begin{tabular*}{\hsize}{@{}@{\extracolsep{\fill}}lccc@{}}
\hline\hline
Parameter                                                               & Value                     & Error      & Unit                 \\\hline
Magnetic tip                                                                                                                            \\
$\quad$ longer diagonal length (part 1) $l_1$                                    & 1.67    	                & 0.05	     & $\mu$m               \\
$\quad$ longer diagonal length (part 2) $l_2$                                    & 4.55     	                & 0.13	     & $\mu$m               \\
$\quad$ half shorter diagonal length $w_s$                              & 2.87    	                & 0.13	     & $\mu$m               \\
$\quad$ tip apex offset $l_0$                                           & 0.98    	                & 0.04	     & $\mu$m               \\
$\quad$ tip height $h_{tip}$                                            & 8.15    	                & 0.27	     & $\mu$m               \\
Cantilever                                                                                                                              \\
$\quad$ length $l$                                                      & 121.90                    & 0.28       & $\mu$m               \\
$\quad$ width $w$                                                       & 34.77                     & 0.47       & $\mu$m               \\
$\quad$ thickness $t$                                                   & 1.62                      & 0.15       & $\mu$m               \\
$\quad$ tilt angle $\theta$                                             & 175.8                     & 0.3        & mrad                 \\
$\quad$ resonance frequency $f_1$                                       & 142.31    	            & 0.14	     & kHz                  \\
$\quad$ quality factor $Q$                                              & 6464                      & 40         &                      \\
$\quad$ spring constant $k$                                             & 3.6                       & 0.4        & N/m                  \\
Nucleon source structure                                                                                                                \\
$\quad$ width of the gold stripe                                        & 9.87                      & 0.06       & $\mu$m               \\
$\quad$ width of the SiO$_{2}$ stripe                                   & 10.23                     & 0.06       & $\mu$m               \\
$\quad$ thickness                                                       & 1.01                      & 0.04       & $\mu$m               \\
\hline\hline
\label{parameter}
\end{tabular*}
\end{table}

\subsection{Displacement amplitude measurement }	
The experiment has been performed on a home-built scanning probe microscope as described in ref. \cite{Wang2016, Ding2020}. The measurements have been conducted in frequency modulation atomic force microscopy (FM-AFM) mode with constant drive, where the cantilever is excited by a piezo plate with a constant drive amplitude. The drive frequency is tracked to the resonance of the cantilever dynamically with phase-locked loop control. The displacement of the cantilever is measured by a fiber interferometer. Its displacement sensitivity,  $S_{int} = \mathrm{d}z/\mathrm{d}V_{int}$, is obtained through polynomial fitting of the cavity length dependent signal $V_{int}$ near the working point where the cavity length is adjusted to maximize the sensitivity. The displacement amplitude is demodulated with a lock-in amplifier. The noise of the amplitude measurement is evaluated at large tip-sample distance where the tip is supposed to be free of sample's action. The standard deviation is estimated to be 14 pm for a acquisition time of 60 ms.

\section{ Results and discussions }

\subsection{ Displacement amplitude imaging }

To avoid accidentally recording data in special location, we have taken two dimensional  (2D) images at several regions of hundreds micrometer away to check the consistency. The images were taken by scanning the magnetic tip over the source mass at constant tip-sample separation. Both oscillation amplitude and interferometer signal were taken simultaneously. The image size is usually 40 $\mu$m $\times$ 40 $\mu$m ($128\times128$ pixels). The fast scan direction is chosen to along the $x$-direction (perpendicular to the stripes). The data acquisition time is 60 ms for every pixel. The sensitivity $S_{int}$ is calculated for every pixel according to its interferometer signal $V_{int}$, which can partially correct the drift of the sensitivity.

\begin{figure}
	\includegraphics [width=8cm]{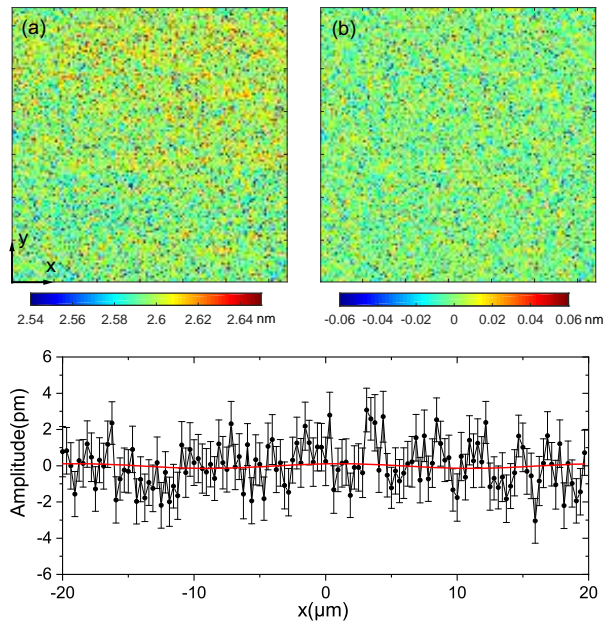}
	\caption{(a) Typical image of the oscillation amplitude taken at a tip-surface distance of 500 nm. (b) Post-processed image after subtraction of the linear background along the $x$-direction line by line. (c) The $x$-dependence of the oscillation amplitude derived from the post-processed image by averaging along the $y$-axis. A simple sine wave fit is also presented. }
	\label{Amplitude_p}
\end{figure}

Figure~\ref{Amplitude_p}(a) shows a typical image taken at a tip-surface distance of 500 nm. The oscillation amplitude is around 2.6 nm. In this image, we observe no obvious periodicity related to the density modulation. In order to highlight the signal variation with respect to the density modulation other than a constant or linear background, we subtract the data by its linear fit along the $x$-direction line by line. The result is presented in Fig.~\ref{Amplitude_p}(b) where random signals are observed with a standard deviation of 14 pm, which is identical to the value obtained at far tip-sample distance. We further average the linearly subtracted data along the $y$-direction to look for the $x$ dependence of the signal. A simple sine wave fit to the data shows that the amplitude is 0.13$\pm$0.17 pm. Such measurements have been performed at four locations laterally separated more than hundreds micrometers away from each other. All images show similar characteristics, and no obvious periodicity has been observed with respect to the density modulation.

The distance dependence of the data can give us important information about the exotic interaction. It is expected that the interaction should decay exponentially with distance. We have performed such measurements at distances of 730 nm, 980 nm, 1294 nm, 1573 nm, 1806 nm.  The images taken at different distances exhibit similar characteristics, and no obvious periodic patterns can be observed. Except that the data points at larger distances unusually have larger values, the signal is almost independent of distance within the fitting error (see Fig.~\ref{Amplitude_d}). The larger values may be due to the unstability of the cantilever oscillation for these two data sets. The result implies that no novel exotic interaction is observed in this experiment. We will then derive the constraints on the coupling constant in the following sections using the data taken at the shortest distance.

\begin{figure}
	\includegraphics [width=8cm]{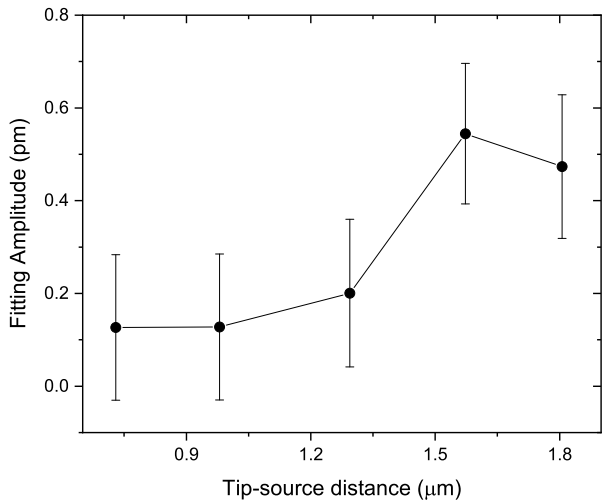}
	\caption{Distance dependence of the signal obtained by fitting the 2D images to the sine wave surface function $z(x, y) = A \mbox{sin}[2\pi(x-x_0)/\Lambda]$, where $\Lambda$ is the density modulation period. }
	\label{Amplitude_d}
\end{figure}

\subsection{ Exotic force calculation }\label{force_cal}

In order to extract the coupling constant of the exotic force from the experimental data, the exotic force is calculated by numerical integration. The interaction potential is obtained by integrating all spins over the coating layer of the tip and all the nucleons in the source mass, which is given by
\begin{widetext}
\begin{equation}
V_{12+13}(x,d)=g_A^eg_V^N\frac{\hbar}{4\pi} \int_{tip} \mathrm{d}S_t \int_{source}\mathrm{d}V_s n(x_s,y_s,z_s)\frac{R_{so}\vec M_r^a \cdot\vec{v}}{\mu _B}\frac{1}{r}e^{-r/\lambda},
\label{eq4}
\end{equation}
\end{widetext}
where $n(x_s,y_s,z_s)$ is the nucleon number density at $(x_s,y_s,z_s)$ in the source mass, $r$ is the distance between the spin on the tip and the source mass element, $\mu_B$ is the Bohr magneton, and $R_{so}$ is the ratio of spin magnetic moment to the total magnetic moment. The angle ($\theta$) between the cantilever and the source surface is considered in the calculation. The potential is a function of the relative lateral position $x$ and distance $d$ between the tip and the source mass. By taking derivative of Eq. (\ref{eq4}) with respect to $d$, the $z$ component of the exotic force is derived. The $R_{so}$  is determined to be 0.92 according to the spin-polarized density functional theory (DFT) calculation of the CoCr alloy (see details in Appendix. A). An example of the expected force signal is shown in Fig.~\ref{Schematic}(b), and we can see that the force is periodic with respect to the relative position $x$.

\subsection{ Spurious forces }
The exotic force depends on the relative velocity, so that the higher oscillation frequency, the greater the expected signal. That is why we chose a stiffer cantilever with a higher resonance frequency. At such high frequency, the vibration noise of the environments is extremely small. For the spurious forces, such as the Casimir force, electrostatic force and magnetic force, they are all independent of the relative velocity. They affect the oscillation amplitude of the cantilever by introducing an extra effective spring constant equal to the force gradient ($k_{ts}=-\partial F_{ts}/\partial z$) between the tip and source mass.

We calculated the Casimir force and electrostatic force by using the proximity force approximation\cite{BORDAG20011}. At a tip-surface distance of 500 nm, the Casimir force gradient is estimated to be $2.5 \times 10^{ - 9}$ N/m, and the electrostatic force gradient is estimated to be  $2.0 \times 10^{- 10}$ N/m with 5 mV residual potential difference between the tip and source mass. Their impact on the oscillation amplitude is less than $10^{ - 8}$ nm, which is negligible compared to the noise floor of the amplitude measurement. The normal magnetic force between the magnetic tip and the source mass is evaluated with finite element simulation. In order to calculate the tiny force, the source mass is simulated as a diamagnetic material of exaggerated magnetic susceptibility, and then the result is extrapolated to the case of gold and silicon dioxide. The calculated force gradient is approximately 8.51 $\times 10^{ - 6} $  N/m on gold, and 8.54 $\times 10^{ - 6}$ N/m on SiO$_{2}$. Their effect on the amplitude is $\sim$ 3  fm (the difference between Au and SiO$_2$ should be smaller), which is also negligible in this experiment. Therefore, the influence of the above forces can be ignored thanks to the velocity dependence of the exotic force and the resonance detection scheme.

\subsection{ Constraints on the coupling constant }

To set constraints on the coupling strength of the exotic interaction, we analyze the images with maximum likelihood estimation. The probability density of observing the image is calculated as a function of the position offset $x_{0}$ and the coupling strength $f_v=2g_A^e g_V^N$ for every $\lambda$ according to
\begin{equation}
P(x_{0},f_v,\lambda)=\frac{1}{A}\prod{\frac{1}{{\sqrt {2\pi }{\sigma _{ij}}}}{e^{-{{\left[{f_{ij}^{\exp} - f_{ij}^T\left({{f_v},\lambda}\right)}\right]}^2}/2\sigma _{ij}^2}}},
\label{eq5}
\end{equation}
where $A$ is the normalization coefficient, $f_{ij}^{\exp }$, $f_{ij}^T\left( {{f_v},\lambda } \right)$ and $\sigma _{ij}$ are the experimental data, theoretical value, and total experimental uncertainty at pixel $(i,j)$, respectively.

The exotic force per unite velocity $f_{ij}^T$ is calculated as described in section \ref{force_cal}. Here, we conservatively use the $2\sigma$ bound values of the experimental parameters to calculate the theoretical values. It should be noted that only spins in the dashed rectangle marked in Fig.~\ref{MFM_Tip}(a) are used in the calculation, so that the real value should be larger. The $f_{ij}^{\exp }$ is the  experimental data at pixel $(i,j)$, converted from the oscillation amplitude image by Eq. (\ref{eq3}). The experimental uncertainty of $f_{ij}^{\exp }$ is the quadrature sum of the errors contributed from $k$, $Q$, $\omega_0$, $z_{far}$ and $z_{amp} (i,j)$. The parameters $\omega_0$, $Q$ and $z_{far}$ were measured at far distance where the cantilever is assumed to vibrate freely. The mean values and uncertainties of those parameters are list in Table~\ref{parameter}. The uncertainty of amplitude measurement $\delta z_{amp}$ is set to 14 pm as evaluated independently at far distance.  An example of the probability density function is shown in Fig.~\ref{Probability_P}.

\begin{figure}
	\includegraphics [width=8cm]{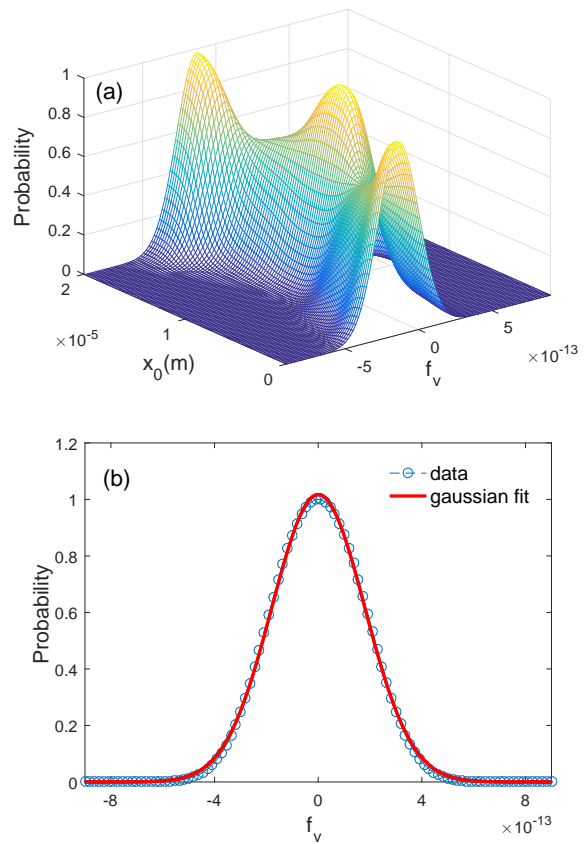}
	\caption{(a) Normalized probability density as a function of $x_{0}$ and $f_v$ for $\lambda$ = 1 $\mu$m . (b) Normalized probability density as a function of $f_v$ by intergrating out $x_0$.The curve is fitted to the Gaussian function,which gives $f_{v} = 3.3 \times 10^{-16} \pm 3.6 \times 10^{-13}$ at 95\% confidential level.}
	\label{Probability_P}
\end{figure}

The probability as a function of $f_v$ is obtained by integrating out the $x_{0}$, and the up-bound of $f_v$ at 95\% confidential level can be derived for every $\lambda$. The result, based on the data taken at 730 nm distance, is presented in Fig.~\ref{Constraints} together with other experimental constrains reported previously.  The atomic magnetometer experiment constrains the exotic force above 180 $\mu$m \cite{Kim2019}.  For the interaction range below 2.3  $\mu$m, the limits are set by comparing the atomic PNC experimental result and theoretical calculations\cite{Dzuba2017}. This work sets stronger constraints than the PNC experiment at the interaction range from 2.3 $\mu$m to  180 $\mu$m. We noticed that a recent experiment using a single-electron spin quantum sensor reported a stronger limit in this range\cite{Jiao2020}. Even stringent constraints may be obtained by the combination of $g_A^e$ and $g_V^N$ from different experiments. We derive the limit on $(g_A^e)^2$ from the spin-spin experiments \cite{ Kotler2015, Rong2018}.  By assuming  $g_s^N =0 $ \cite{Fadeev2019, Raffelt2012}, the limit on $(g_V^N)^2$ can be obtained from the non-Newtonian gravity experiments \cite{Chen2016, Tan2020, Lee2020}.  Finally, we obtain the limit on the combination $g_A^eg_V^N$, as shown by the dashed line  in Fig. 7.  Nevertheless, direct experimental search with fewer assumptions are still worthwhile.

\begin{figure}
	\includegraphics [width=8cm]{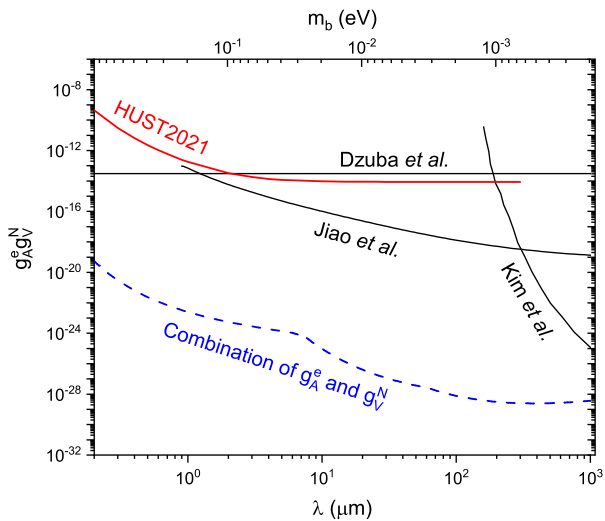}
	\caption{Constraints on the dimensionless coupling constants $g_A^e g_V^N$  from this work as well as previous experiments\cite{Dzuba2017,Kim2019,Jiao2020}. The dashed line shows the limit on the combination of $g_A^e$ and $g_V^N$ as explained in the main text. }
	\label{Constraints}
\end{figure}

\section{ Conclusions }

 In conclusion, we have performed an experimental search for the exotic parity-odd spin- and velocity-dependent interaction with a magnetic force microscope. The magnetic probe is taken as a source of the moving spin polarized electrons and also the force sensor to measure the exotic interaction. The magnetic properties of the tip is evaluated quantitatively by VSM and qualitatively by MFM measurement on a hard-disk plate. The density modulation source mass is adopted to generate a periodic signal, so that we are able to extract the exotic interaction from non-periodic background. We obtain the coupling constant by using maximum likelihood estimation, and observe no new interaction above the experimental sensitivity. The experiment sets constraints on the coupling constant between spin polarized electrons and unpolarized nucleons at the micrometer range using a direct force measurement method. Future improvements could be achieved by replacing the magnetic tip with a spherical magnet. The spherical magnet of micrometer size can be magnetized to single domain due to its symmetry and could provide more spins to interact with the source mass.


\begin{acknowledgments}
The authors thank Fei Xue for useful discussions. This work was supported by the National Natural Science Foundation of China under Contract Nos. 11875137, 91736312, 91436212, 51972312 and 51472249. The DFT calculations in this work were performed on TianHe-1(A) at the National Supercomputer Center in Tianjin.

\end{acknowledgments}

\appendix
\section{Density functional theory calculation}
The spin-polarized DFT calculations are performed by using the Vienna Ab Initio Simulation Package (VASP) code\cite{Kresse1996}. Perdew-Burke-Ernzerhof (PBE) functional is employed to describe the exchange correlation interactions within the generalized gradient approximation. The electron-ion interactions are represented by the projector augmented wave (PAW) method\cite{Perdew1996, blochl1994}. The energy cutoff of plane wave is set to be 500 eV and the convergence criterion for the residual forces and total energies are set to be 0.01 eV/$\r{A}$ and $10^{-6}$ eV, respectively. The primitive cell with two Co atoms, a 2$\times$2$\times$1 supercell with 6 Co and 2 Cr atoms, and a 2$\times$2$\times$2 supercell with 13 Co and 3 Cr atoms are constructed to calculate the spin and orbital magnetic moment of $hcp$-Co, $hcp$-Co$_{3}$Cr and $hcp$-Co$_{13}$Cr$_3$ alloy, respectively. The $\Gamma$-centered 9$\times$9$\times$6, 5$\times$5$\times$6 and 5$\times$5$\times$2 Monkhorst-Pack k-point mesh are used to sample the Brillouin zones of $hcp$-Co, $hcp$-Co$_{3}$Cr and $hcp$-Co$_{13}$Cr$_3$ alloy, respectively. The $R_{so}$ values are found to be  0.96, 0.93, 0.90 for pure Co, Co$_{13}$Cr$_3$  and Co$_{3}$Cr, respectively. The value decreases with increasing the concentration of Cr. By interpolating the data, we estimate the $R_{so}$ value for Co$_{80}$Cr$_{20}$ to be 0.92.

%


\begin{thebibliography}{71}
\bibitem [{MFM()}]{MFMtip00}%
  \BibitemOpen
  \href@noop {} {}\bibinfo {note} {Bruker MFM Probes - MESP-RC-V2, it has a
  nominal coercivity of 400 Oe (medium), and a magnetic moment of 1e-13 emu
  (medium)}\BibitemShut {NoStop}%
\end{thebibliography}
\end{document}